# A New Design Technique of Reversible BCD Adder Based on NMOS With Pass Transistor Gates

Md. Sazzad Hossain[1], Md. Rashedul Hasan Rakib[1], Md. Motiur Rahman[1], A. S. M. Delowar Hossain[1] and Md. Minul Hasan[2]

[1]Department of Computer Science and Engineering, Mawlana Bhashani Science & Technology University, Santosh, Tangail-1902, Bangladesh
sazzad_101@yahoo.com
[2]Amader Ltd, 5B Union Erin, 9/1 North Dhanmondi, Kalabagan, Dhaka, Bangladesh.
md.minul.hasan@gmail.com

## ABSTRACT

*In this paper, we have proposed a new design technique of BCD Adder using newly constructed reversible gates are based on NMOS with pass transistor gates, where the conventional reversible gates are based on CMOS with transmission gates. We also compare the proposed reversible gates with the conventional CMOS reversible gates which show that the required number of Transistors is significantly reduced.*

## 1. INTRODUCTION

Irreversible hardware computation results in energy dissipation due to information loss. According to Landauer's research, the amount of energy dissipated for every irreversible bit operation is at least KTln2 joules, where K=1.3806505*10-23m2kgs-2K-1 (joule/kelvin) is the Boltzmann's constant and T is the temperature at which operation is performed [1, 2]. In 1973, Bennett showed that KTln2 energy would not dissipate from a system as long as the system allows the reproduction of the inputs from observed outputs [3, 4].

Reversible logic supports the process of running the system both forward and backward. This means that reversible computations can generate inputs from outputs and can stop and go back to any point in the computation history. Thus, reversible logic circuits offer an alternative that allows computation with arbitrarily small energy dissipation. Therefore, logical reversibility is a necessary (although not sufficient) condition for physical reversibility.

There are many design techniques to implement a reversible BCD adder. The internal reversible gates of those BCD adders are used CMOS pass transistor logic [5, 6]. In this paper, we have avoided conventional CMOS based reversible gates and implement those reversible gates using NMOS based pass transistor logic. Finally, we have proposed a reversible BCD adder that poses all the good features of reversible logic synthesis using our implemented reversible gates.

The rest of the paper is composed of a number of sections. Section two- Background; describes the origin of various reversible gates. Section three-Properties of Pass Transistor; describes the attributes and general operations of pass transistor. Section four-Construction of proposed reversible gates; describes how to construct our proposed reversible gates from the conventional reversible gates. Section five and six- Design of a Reversible Full Adder and BCD adder;

describes how to construct a Reversible Full Adder and BCD Adder using our proposed reversible logic gates. Section seven-Comparison; describes the performance of our proposed technique. Conclusion has been drawn in the Last Section.

## 2. BACKGROUND

In conventional (irreversible) circuit synthesis, one typically starts with a universal gate library and some specification of a Boolean function. It is widely known that an arbitrary Boolean function can be implemented using only NAND gates. A NAND gate has two binary inputs (say A, B) but only one binary output (say P), and therefore is logically irreversible.

### 2.1. Reversible Gates and Circuits

Fredkin and Toffoli have shown in [8] that a basic building block which is logically reversible should have three binary inputs (say A, B and C) and three binary outputs (say P, Q and R).

Feynman has proposed in [1], [9] the use of three fundamental gates:
- The NOT gate,
- The CONTROLLED NOT gate and
- The CONTROLLED CONTROLLED NOT gate.

Together they form a set of three building blocks with which we can synthesize arbitrary logic functions.

The NOT gate can be realized,
P=NOT A

The CONTROLLED NOT can be realized,

When P=A and
If   A=0,  then Q=B
         else Q=NOT B
So we can write Q=A XOR B

Table 1: Truth table of CONTROLLED NOT

| A | B | P | Q |
|---|---|---|---|
| 0 | 0 | 0 | 0 |
| 0 | 1 | 0 | 1 |
| 1 | 0 | 1 | 1 |
| 1 | 1 | 1 | 0 |

The CONTROLLED CONTROLLED NOT can be realized,

When P=A, Q=B and

If A AND B=0, then R=C,
         Else R=NOT C

So we can write R= (A AND B) XOR C

Table 2: Truth table of CONTROLLED CONTROLLED NOT

| A | B | C | P | Q | R |
|---|---|---|---|---|---|
| 0 | 0 | 0 | 0 | 0 | 0 |
| 0 | 0 | 1 | 0 | 0 | 1 |
| 0 | 1 | 0 | 0 | 1 | 0 |
| 0 | 1 | 1 | 0 | 1 | 1 |
| 1 | 0 | 0 | 1 | 0 | 0 |
| 1 | 0 | 1 | 1 | 0 | 1 |
| 1 | 1 | 0 | 1 | 1 | 1 |
| 1 | 1 | 1 | 1 | 1 | 0 |

The CONTROLLED CONTROLLED NOT has a significant characteristic: it is a universal primitive which means, by combining a finite number of such building blocks, any Boolean function of any finite number of logic input variables can be implemented.

FREDKIN gate also possesses the characteristic of the CONTROLLED CONTROLLED NOT that is it is another universal primitive. It can be realized,

When P=A and
If A=0, then Q=B
             R=C
        else Q=C
             R=B

So we can write Q= ((NOT A) AND B) OR (A AND C)
                R= ((NOT A) AND C) OR (A AND B)

Table 3: Truth table of Fredkin gate

| A | B | C | P | Q | R |
|---|---|---|---|---|---|
| 0 | 0 | 0 | 0 | 0 | 0 |
| 0 | 0 | 1 | 0 | 0 | 1 |
| 0 | 1 | 0 | 0 | 1 | 0 |
| 0 | 1 | 1 | 0 | 1 | 1 |
| 1 | 0 | 0 | 1 | 0 | 0 |
| 1 | 0 | 1 | 1 | 1 | 0 |
| 1 | 1 | 0 | 1 | 0 | 1 |
| 1 | 1 | 1 | 1 | 1 | 1 |

## 3. PROPERTIES OF PASS TRANSISTOR

### 3.1. Pass Transistor Logic

Pass transistor NMOS based transistor which has a control signal P1. The control signal P1 is responsible for transferring the input signal V1 (pass signal) to the output [10]. This works like a switching circuit. When the P1 is activated, then the Input signal V1 will pass through the gate and will go to the Output. But the Input signal cannot pass without the activation of Control signal P1.

Table 4: The truth table for pass transistor logic is as follows:

| Control Signal (P1) | Input Signal (V1) | Output |
|---|---|---|
| 0 | 0 | High Impedance |
| 0 | 1 | High Impedance |
| 1 | 0 | 0 |
| 1 | 1 | 1 |

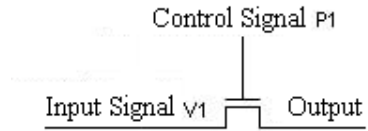

Figure 1: Model for Pass transistor logic

### 3.2. Threshold Voltage

A pass transistor with a threshold gate is shown in Figure 2. The threshold gate is replaced by a conventional NMOS or CMOS inverter.

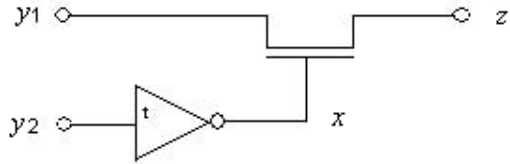

Figure 2: Symbol of a pass transistor and a threshold gate

If the pass transistor in figure 2 is turned on, the output is equal to the input, while if it is turned off, then the output is in a high impedance state.

The inverting voltage $V_i$ is represented for an NMOS inverter as:

$$v_i = -k v_d + v_e \quad (1)$$

and for an CMOS inverter as:

$$v_i = (k(v_{dd} + v_{tp}) + v_{tn})/(k+1) \quad (2)$$

Each threshold voltage $v_t$ in MOS transistors can be fabricated by the ion-injection technology with high accuracy. Therefore, those inverters can be used as inverted threshold gates with arbitrary threshold values.

### 3.3. Representation of pass transistor

Figure 2 shows the symbolic representations for a pass transistor with an inverted threshold gate. The definition of the values in the inputs $y_1$ and $y_2$ and the output $z$ is given as used in [11]:

$$y_r, z \in (0, 1, 2 \ldots r-1, \Phi) \quad (3)$$

The relation between $y_2$ and $x$ in the inverted threshold gate is defined as:

$$x = \begin{cases} 0 & \text{for} \quad y_2 \geq t \\ 1 & \text{for} \quad y_2 < t \end{cases} \quad (4)$$

Using the internal parameter $x$, the relation of the input $y_1$ and the output $z$ in a pass transistor is denoted as:

$$z = y_1 < x > = \begin{cases} y_1 & \text{for} \quad x = 1 \\ \emptyset & \text{for} \quad x = 0 \end{cases} \quad (5)$$

### 3.4. Connection of pass transistor

As described in detailed in [10] the pass transistors with threshold gates can be combined in series and/or parallel connection combinations. The equation (5) can be regarded as the basic of the representation of the inputs and outputs of connections [13].

### 3.4.1. Series connection:

The series connection can be depicted as:

$$z = y < x_1, x_2 \ldots x_n > \quad (6)$$

This is shown in Figure 3

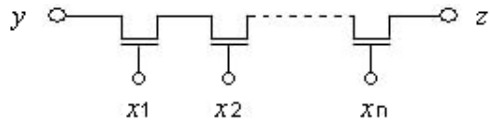

Figure 3: Series connection

### 3.4.2. Parallel connection:

Parallel connections for common inputs can be depicted by the equation:

$$z = y < x_1 \vee x_2 \vee \ldots \vee x_n > \quad (7)$$

This is shown in Figure 4.

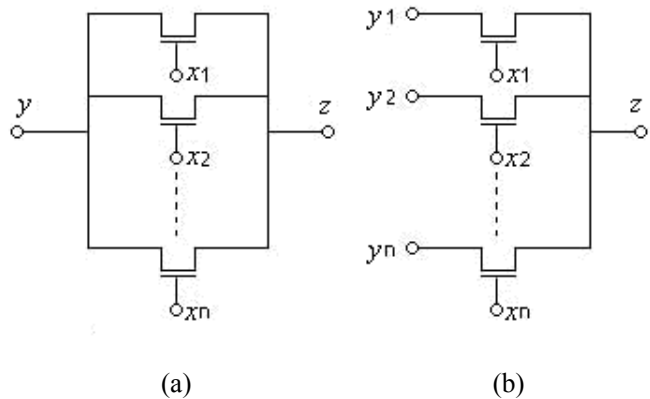

(a)          (b)

Figure 4: Parallel connection a) Common inputs b) Different inputs

While the parallel connections for different inputs can be depicted as:

$$z = y_1 < x_1 > + y_2 < x_2 > + \ldots + y_n < x_n > \quad (8)$$

Where '+' is an operator which satisfies the Commutative law, Law of absorption, Distributive law etc.

## 3.5. Parallel-series connection

The combinations of series-parallel connections can be depicted as, for common inputs (figure 5).

$$y_2 = y_1 <x_1 \vee x_2>$$
$$\begin{aligned}z &= y_2 <x_3>\\ &= (y_1 <x_1 \vee x_2>) <x_3>\\ &= y_1 <(x_1 \vee x_2)x_3>\end{aligned} \quad (9)$$

And for different inputs (Figure 5).

$$y_3 = y_1 <x_1> + y_2 <x_2>$$
$$\begin{aligned}z &= y_3 <x_3>\\ &= (y_1 <x_1> + y_2 <x_2>) <x_3>\\ &= y_1 <x_1 x_3> + y_2 <x_2 x_3>\end{aligned} \quad (10)$$

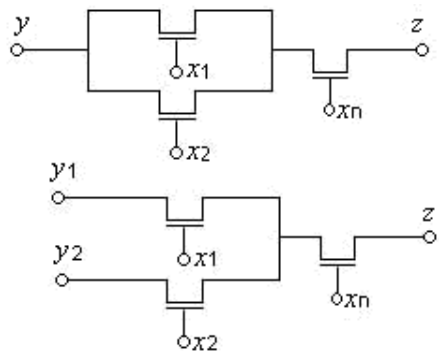

Figure 5: Parallel – series connection (Common Inputs and Different inputs)

## 4. CONSTRUCTION OF PROPOSED REVERSIBLE GATES

### 4.1. NOT

The construction of the reversible NOT gate does not require any transistor [1]. That is why it is the simplest form of reversible gate.

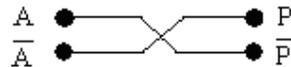

Figure 6: Reversible NOT gate

### 4.2. CONTROLLED NOT

The CONTROLLED NOT reversible gate has two types of inputs. The first one is control inputs (A, $\overline{A}$) they control the ON and OFF states of the transistors and thus control the transfer of pass signals (B, $\overline{B}$) from input to output.

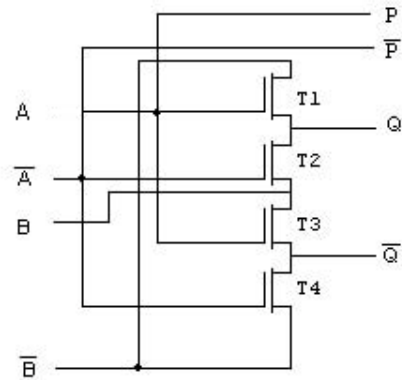

Figure 7: Reversible CONTROLLED NOT gate

Now consider a set of values for control and pass signals to realize the operation of CONTROLLED NOT gate. Say, A=1, B=1 And corresponding inverted inputs are $\bar{A}$=0, $\bar{B}$=0
The input A controls the transistors T1 and T3 and $\bar{A}$ controls the transistors T2 and T4. For this control the T1 and T3 are ON and T2 and T4 are OFF and we get the pass signal (B, $\bar{B}$) in the corresponding output lines P=1, $\bar{P}$=0 and Q=0, $\bar{Q}$=1.

Now if we consider the reverse mode of operation then control inputs will be P, $\bar{P}$ and pass signals will be Q, $\bar{Q}$ and outputs will be A, $\bar{A}$ and B, $\bar{B}$. So if we start with the outputs of the forward mode of operation as inputs in the reverse mode we obtain the inputs of the forward mode as the outputs of the reverse mode.

### 4.3. CONTROLLED CONTROLLED NOT

The CONTROLLED CONTROLLED NOT reversible gate is similar in operation with respect to CONTROLLED NOT; Except that controlling of transistors occurs twice while transferring pass signal from input to output.

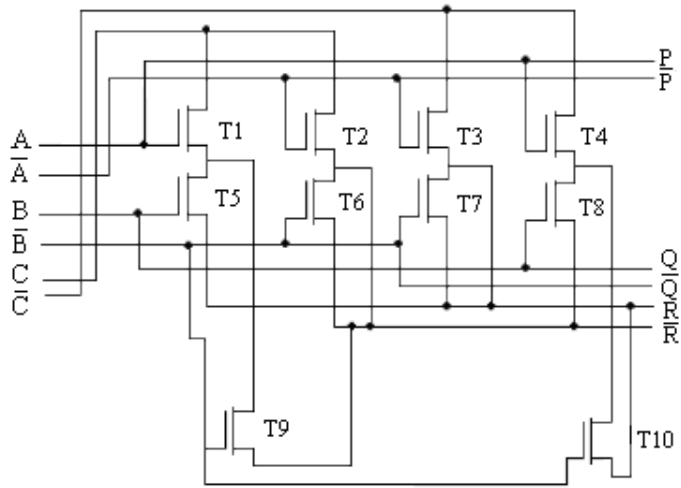

Figure 8: Reversible CONTROLLED CONTROLLED NOT gate

As the control mechanism occurs twice in this reversible gate that is why this gate is referred to as CONTROLLED CONTROLLED NOT.

### 4.4. FREDKIN GATE

In FREDKIN gate the control mechanism occurs only once by the control input A while transferring the pass signals B and C to the outputs P, Q and R.

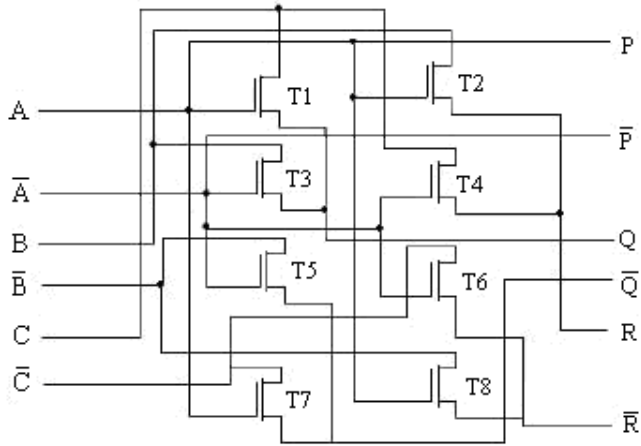

Figure 9: Reversible FREDKIN gate

## 5. DESIGN OF A REVERSIBLE FULL ADDER

If we consider a conventional Full Adder [1], the input section generally contains two inputs A and B with another input which is known as the carry in ($C_i$). The output section contains two outputs Carry out ($C_o$) and Sum (S). For different combination of input values, the conventional Full Adder sometimes generates the same outputs. That is why; we cannot determine the specific input values by looking at the output values. This yields that the circuit cannot be used as a reversible circuit. The solution of this problem is to make each of the output set unique, so that by judging the output we can determine the corresponding input. Thus the Adder can be operated reversibly.

To achieve this uniqueness of the output values we need to add some extra bits both in input and output section. The extra bit in the input section is Preset input (P) and the extra bits in the output section are two garbage outputs $G_1$ and $G_2$ respectively. The truth table for reversible Full Adder has shown in Table 5. The implementation of Full Adder using reversible logic is as follows:

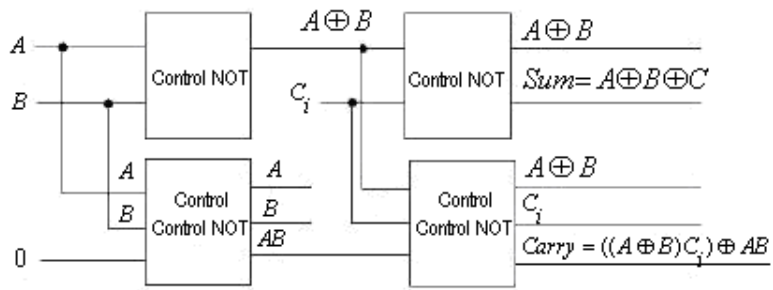

Figure 10: Reversible Full Adder with two CONTROLLED NOT gates and two CONTROLLED- CONTROLLED NOT gates.

Table 5: Truth table for reversible Full Adder

| A | B | $C_i$ | P | SUM | CARRY | $G_1$ | $G_2$ |
|---|---|---|---|---|---|---|---|
| 0 | 0 | 0 | 0 | 0 | 0 | 0 | 0 |
| 0 | 0 | 0 | 1 | 1 | 0 | 0 | 0 |
| 0 | 0 | 1 | 0 | 0 | 1 | 0 | 0 |
| 0 | 0 | 1 | 1 | 1 | 1 | 0 | 0 |
| 0 | 1 | 0 | 0 | 0 | 1 | 0 | 1 |
| 0 | 1 | 0 | 1 | 1 | 1 | 0 | 1 |
| 0 | 1 | 1 | 0 | 1 | 0 | 0 | 1 |
| 0 | 1 | 1 | 1 | 0 | 0 | 0 | 1 |
| 1 | 0 | 0 | 0 | 0 | 1 | 1 | 1 |
| 1 | 0 | 0 | 1 | 1 | 1 | 1 | 1 |
| 1 | 0 | 1 | 0 | 1 | 0 | 1 | 1 |
| 1 | 0 | 1 | 1 | 0 | 0 | 1 | 1 |
| 1 | 1 | 0 | 0 | 1 | 0 | 1 | 0 |
| 1 | 1 | 0 | 1 | 0 | 0 | 1 | 0 |
| 1 | 1 | 1 | 0 | 1 | 1 | 1 | 0 |
| 1 | 1 | 1 | 1 | 0 | 1 | 1 | 0 |

It can be easily verified from the above table that it fulfils two conditions: (a) all sixteen outputs ($C_o SG_1 G_2$) are different, such that the table is reversible, and (b) if P=0, than the output column $C_o$ and S have the values of a conventional Full Adder. The outputs ($C_o SG_1 G_2$) contain the same amount of information as the inputs ($ABC_i P$), such that backward calculation is possible.

The construction of the reversible Full Adder consists of our proposed CONTROLED NOT and CONTROLED -CONTROLED NOT reversible logic gates.

## 6. PROPOSED BCD ADDER USING REVERSIBLE FULL ADDER

For Designing a BCD adder we need to Design a reversible circuit for the expression C3 + S3 (S1 + S2). Where C3 denotes the carry out of the top reversible adder and S3, S2 and S1 is the sum bits of the top adder shown in figure 12. For the above expression we design a new reversible gate using pass transistor logic for calculating OR operation shown in Table 6. We have said this gate as New Proposed Gate (NPG). The logic for this gate can be realized:

When P=A and
If A=1 then Q = A
    Else Q = B
So we can write Q = A + B

The truth table for this proposed gate is as follows:

Table 6: Truth table for new reversible Gate

| A | B | P | Q |
|---|---|---|---|
| 0 | 0 | 0 | 0 |
| 0 | 1 | 0 | 1 |
| 1 | 0 | 1 | 1 |
| 1 | 1 | 1 | 1 |

The proposed circuit for the above table is as follows:

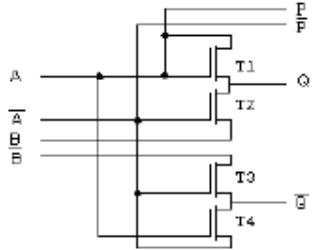

Figure. 11 New Proposed Reversible gate

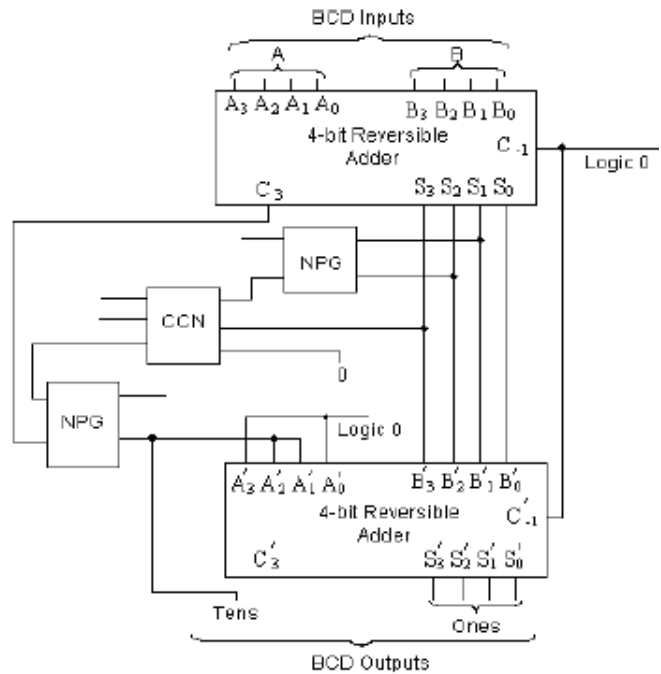

Figure. 12 BCD adder circuit using reversible gates

## 6. COMPARISON

Pass transistor can be constructed by using either NMOS or CMOS. The NMOS version has the fastest fall time and the CMOS version has the fastest rise time, but requires a pre-charge period that may extend clock cycle time. Complementary pass networks incur extra pull down delay. In comparison with regular gates, the merging of source and drain regions is difficult, leading to higher internal node capacitances. The CMOS construction requires full transmission gates and if we compare the number of transmission gates with the NMOS construction, we find that NMOS requires fewer transistors. Pass networks, due to their lower stray capacitance are good for low power and high performance systems. The following table describes the exact amount of transistors required to construct reversible gates using proposed NMOS based technique and conventional CMOS based technique.

Table 7: No. of transistors used in proposed NMOS and Conventional CMOS based reversible gates

| Gates | Proposed NMOS | Conventional CMOS |
|---|---|---|
| NOT | 0 | 0 |
| CONTROLLED NOT | 4 | 8 |
| CONTROLLED-CONTROLLED NOT | 10 | 16 |
| FREDKIN | 8 | 16 |

Using the conventional CMOS reversible transmission gates, the construction of the circuit described in figure 10 will require 2 x 8 + 2 x 16 = 48 transistors. Meanwhile, the construction of the same circuit using the proposed reversible gates will require 2 x 4 + 2 x 10 = 28 transistors. So our proposed design significantly reduces the number of transistors in the circuit. Our proposed BCD adder requires 2 x 28 + 10 + 4 x 2 = 74, which is approximately 50% less than the conventional reversible BCD adder.

## 7. CONCLUSIONS

In this paper an optimized reversible BCD adder is presented using our proposed reversible gates. The design is very useful for the future computing techniques like low power digital circuits and quantum computers. We have constructed the NMOS based reversible gates. It uses fewer transistors then conventional CMOS based reversible gates. This yields that the amount of space of the circuitry is reduced, as well as the gates operate with low power dissipation. The time of operation should also decrease i.e. the circuit will work much faster. An important characteristic of the circuit is that all energy supplied to the system is delivered by the input signals themselves. The design method is definitely useful for the construction of future computer and other computational structures. The proposed technique can be enhanced in future by testing this proposed design in various simulators and in real world application circuits.

**Authors**

**Md. Sazzad Hossain** received his B.Sc. (Engg.) in Computer Science & Engineering from Mawlana Bhashani Science and Technology University, Bangladesh, in 2008. Now he is a faculty member of CSE department of that university.

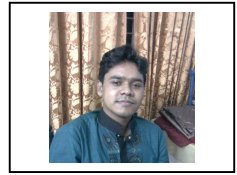

**Md. Rashedul Hasan Rakib** received his B.Sc. (Engg.) in Computer Science & Engineering from Khulna University, Bangladesh, in 2008. Now he is a faculty member of CSE department of Mawlana Bhashani Science and Technology University, Bangladesh.

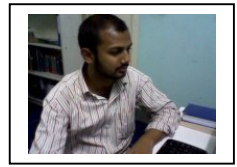

**Md. Motiur Rahman** received his B.Sc. (Engg.) and M.Sc. (Engg.) in Computer Science & Engineering from Jahangirnagar University, Bangladesh. Now he is a faculty member of CSE department of Mawlana Bhashani Science and Technology University, Bangladesh.

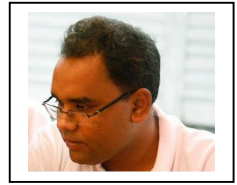

**A. S. M. Delowar Hossain** received his B.Sc. (Engg.) in Computer Science & Engineering from Islamic University of Technology, Bangladesh, in 2008. Now he is a faculty member of CSE department of Mawlana Bhashani Science and Technology University, Bangladesh.

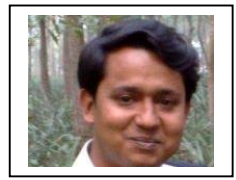

**Md. Minul Hasan** received his B.Sc. (Engg.) in Computer Science & Engineering from Mawlana Bhashani Science and Technology University, Bangladesh, in 2008. Now he is the system administrator of Amader Ltd, Dhaka, Bangladesh.

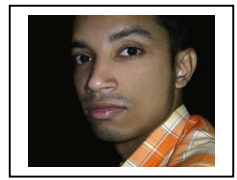